\documentclass[seceq]{jpsj2}

\makeatletter
\def\@citess#1{\textsuperscript{#1)}}
\newcommand{\lbar}{\left|}
\newcommand{\rbar}{\right|}
\newcommand{\bra}{\left\langle}
\newcommand{\ket}{\right\rangle}
\newcommand{\mathvec}[1]{\mbox{\boldmath $#1$}}
\newcommand{\textvec}[1]{{\boldmath $#1$}}
\title{
Spin-Reorientation Transition of Field-Induced Magnetic Ordering Phases in the Anisotropic Haldane System
}
\author{
Hiroshi \textsc{Miyazaki}\thanks{E-mail:miyazaki@science.okayama-u.ac.jp},
Tomohiko \textsc{Hiwasa},
Masataka \textsc{Oko},
Isao \textsc{Harada}
}
\inst{
The Graduate School of Natural Science and Technology, Okayama University, 3-1-1 Tsushima-naka, Okayama 700-8530
}
\abst{
A possible spin-reorientation transition in field-induced magnetic ordering phases of the $S=1$ Haldane system with large easy-plane anisotropy is proposed, using an effective Lagrangian formalism as well as the density matrix renormalization group method.
Such a spin-reorientation transition is predicted in the case where the applied magnetic field is inclined from the easy axis of the anisotropy.
We point out that this transition has a close connection with a variation of the order parameter even at zero temperature, although it is different from a quantum analog of the so-called spin-flop transition proposed for the system having a strong easy axis anisotropy.
In connection with a novel phase observed recently in the Haldane system at high fields, we discuss possible implications for the field-induced magnetic ordering.
}
\kword{
anisotropic Haldane chain,
field-induced phase transition,
spin-reorientation
}
\begin{document}
\sloppy
\maketitle
%
%
\section{Introduction}
Quantum fluctuations exhibit interesting roles in one-dimensional antiferromagnetic spin systems and eventually destroy the antiferromagnetic ordering even at zero temperature.
As a consequence, the ground state turns out to be a novel singlet state having an energy gap to triplet states.
An $S=1$ spin chain with nearest neighbor exchange interactions, the so-called Haldane chain, is a typical example of such systems.\cite{rf:Haldane}

In order to understand such a ground state, many researchers are interested in the recovery of magnetism, magnetic ordering, destroying the gap by a strong external magnetic field, for instance in Ni(C$_2$H$_8$N$_2$)$_2$NO$_2$(ClO$_4$).\cite{rf:Koba}
Honda \textit{et al.} were the first to realize experimentally the field-induced long-range N\'eel ordering at about 4~T in the Haldane system, Ni(C$_{5}$H$_{14}$N$_{2}$)$_{2}$N$_{3}$(PF$_{6}$), abbreviated hereafter as NDMAP.\cite{rf:HAK,rf:HKNH,rf:CHZ,rf:Sakai-pd}
It is noted that the critical behavior of the phase transition depends crucially on the direction of  the magnetic field, suggesting an important role of the anisotropies of a Ni$^{2+}$ spin: it has a rather large easy-plane anisotropy with a weak in-plane anisotropy.\cite{rf:H}

Following the work of Honda \textit{et al.}, Tsujii \textit{et al.} extended the measurement of the heat capacity in fields up to 30~T along the crystallographic $c$ axis and found a new phase transition at 14~T in addition to the former at 4~T.\cite{rf:T}
At first glance, this transition seems to be a quantum analog of the so-called spin-flop transition, which occurs in the case where the ground state is the N\'eel state.\cite{rf:Sakai}
We, however, realize immediately that it is not the case because NDMAP has the disordered ground state and the field-induced ordered moment is perpendicular to it.

Now, we remember the tilting of the principal axes of the anisotropy in NDMAP: the local principal $z$ axis does not coincide with the crystallographic $c$ axis but tilts from the $c$ axis by 16 degrees in the $a$-$c$ plane.
Furthermore, we note that there are two types of chains, each of which belongs to one of the sublattices of the body-centered-cubic lattice and has the tilt direction just opposite to another.
Thus, the antiferromagnetic ordering of the two sublattices are magnetically decoupled at the mean-field level due to the geometric frustration.
It is worth noticing that such an interesting situation has skillfully been confirmed recently by the neutron elastic scattering: a half-ordered state has been observed in NDMAP between the two critical fields for each sublattice, adjusting the field direction to the $z$ axis of one of the two kinds of sublattices.\cite{rf:Z}

In order to understand what happens in NDMAP at high fields, we will consider the following Hamiltonian describing the Haldane system with the planar anisotropy as well as a weak in-plane anisotropy under the external magnetic field,
\begin{equation}
  \label{eq:Hamiltonian}
  \mathcal{H} / J = \sum_i
  \left[\mathvec{S}_i \cdot \mathvec{S}_{i+1} + d (S_i^z)^2
    - e \{(S_i^x)^2-(S_i^y)^2\}
    - \mathvec{h}\cdot\mathvec{S}_i \right],
\end{equation}
where \textvec{S} is the $S=1$ spin operator and the Cartesian coordinates $(x,y,z)$ are referred to the spin anisotropy axes.
The parameters, $J$, $d$, $e$ and \textvec{h} denote, respectively, the exchange constant, which defines the unit of energy throughout this paper, the easy-plane anisotropy constant $d(=0.25)$, the in-plane anisotropy constant $e(=0.01)$ and the external magnetic field.\cite{rf:ZHC,rf:KMMKIY,rf:HKHT}
Note that the anisotropies $d$ and $e$ make the $x$ axis easy while the $z$ axis hard.
Considering that the anisotropy axis in NDMAP tilts in the $a$-$c$ plane,\cite{rf:Z} we assume \textvec{h} has the $x$ and the $z$ components: $\mathvec{h}=h(\sin\theta,0,\cos\theta)$.
A similar system has been studied by the perturbative calculation.\cite{rf:Gol}
The aim of this paper is to explore the role of quantum fluctuations in the field-induced {\em ordering phase} from the unified point of view.
Especially, we will be concerned with the nature of various ordering phases and phase transitions between them, focusing our attention to the field direction.
We are interested in the field region above the critical field $h_\mathrm{c}$, at which the Haldane gap is closed.
To this end, we adopt the density matrix renormalization group (DMRG) method\cite{rf:S} as well as the phenomenological field theory (PFT);\cite{rf:A}
the former is used for \textit{ab initio} simulations for a single spin chain while the latter is suitable for unified understanding of the phenomena.
The former is also used for the confirmation of the results of PFT.
With the aid of the information on the single Haldane chain, we analyze the phase diagram, using the mean-field approximation for weak inter-chain interactions.
It has been confirmed that such an approximation reproduces the phase diagrams of NDMAP quite well.\cite{rf:HKNH}
It is worth noticing that the important information on the criticality of the single spin chain can be drawn from the analysis of the phase diagram.\cite{rf:HKNH,rf:NH}

The present paper is organized as follows.
In the next section, we describe the methods used to obtain the field-induced ordering states in the Haldane system and their excited states.
In \S 3, we give theoretical results and consider the phase diagram and other thermodynamic quantities.
In the last section, we discuss possible implications and relevance of the present results in connection with the experimental results in NDMAP and conclude our discussion with some remarks.

\section{Methods}
In this paper, we perform simulations for the Haldane phase and ordering phases in the anisotropic Haldane chain, using (1) the \textit{ab initio} numerical calculation by the DMRG method, and (2) the Landau-Ginsburg type theory of PFT, within the mean-field approximation.
As will be seen soon later, the direction of the applied magnetic field with respect to the anisotropy axes plays important roles in the physics of the field-induced phase transition of the anisotropic Haldane chain.

In the ordering phase, we define the sublattice magnetization
\begin{equation}
  \mathvec{L} = \bra \mathvec{S}_{2n} - \mathvec{S}_{2n+1} \ket ,
\end{equation}
and characterize possible ordering phases in the $h$-$\theta$ phase diagram in addition to the Haldane phase:
if \textvec{L} is in the $x$-$z$ plane, we call it the $xz$-phase and if \textvec{L} is in the $y$ direction, we call it the $y$-phase.

\subsection{Density matrix renormalization group method}
The DMRG method\cite{rf:S} has been developed in order to reveal properties of one-dimensional strongly correlated systems.
It is known that this method provides the ground state energy and low-lying excited state energies of spin-gaped systems with very high accuracy.
Then, we use the DMRG method to calculate the physical quantities of the Haldane system.
Further, we extend it so that it is applicable to the ordering states, and discuss the phase transition between the $xz$-phase and the $y$-phase.

\begin{figure}[t]
  \centering
  \includegraphics{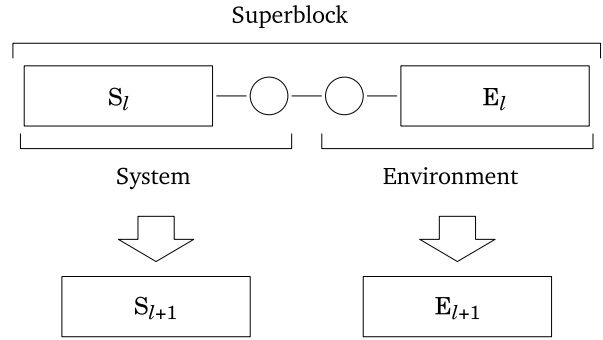}
  \caption{A schematic diagram of the superblock. The superblock of length $2l+2$ is constructed from the system and the environment blocks of length $l$ and two single sites. Blocks of length $l+1$, $\mathrm{S}_{l+1}$ and $\mathrm{E}_{l+1}$, are obtained from the superblock of length $2l+2$.}
  \label{fig:dmrg}
\end{figure}

The anisotropic Haldane chain described by the Hamiltonian (\ref{eq:Hamiltonian}) is studied by the infinite-system DMRG algorithm.\cite{rf:S}
In this algorithm, the whole system is divided into two blocks, \textit{system} and \textit{environment}, as is schematically shown in Fig.~\ref{fig:dmrg}.
These blocks are elongated iteratively keeping the Hilbert space up to a certain size $m$.
The bases of the system block of length $l+1$ are chosen from eigenstates with the largest eigenvalue of the reduced density matrix:
\begin{equation}
  \label{eq:rho}
  \rho^{\mathrm{S}} = \mathrm{Tr_E} \lbar \psi \ket \bra \psi \rbar,
\end{equation}
where $\mathrm{Tr_E}$ means tracing out the degrees of freedom of environment side and $\psi$ is the ground state wave function of the \textit{superblock} of length $2l+2$, which is built from the system and environment blocks of length $l$ and two single sites between them.
Usually, we can form the environment block by reflecting the system block.

The calculation has been performed under the open boundary condition (OBC).
As is well known, since the Haldane system with OBC has two effective $S=1/2$ degrees of freedom at each end of the chain.\cite{rf:WH}
Then, we add a real $S=1/2$ spin at each end in order to remove the ground state degeneracy.

In order to study the ordering phase, we need to extend DMRG so that it is able to treat a phase in which the $\mathrm{Z}_2$ symmetry is broken.
To this end, we first apply a weak magnetic field to an edge spin of the superblock to break the $\mathrm{Z}_2$ symmetry.
As a consequence, we choose the ground state $\lbar \psi \ket$ of the superblock uniquely.
Then, we construct the system block and the environment block, using eq.~(\ref{eq:rho}) individually for each block.
Repeating these procedures, we finally obtain the ordering state.

We compute the energy gap $\Delta$, which is obtained by subtracting the ground state energy of the superblock from the lowest excited state energy, and the sublattice magnetization \textvec{L} in the ordering phase, which is evaluated at two spins located on the center of the superblock, as a function of the field $h$ at various values of the field angle $\theta$.
As the infinite-system DMRG is often trapped into metastable states, the calculation has been done starting both from the $xz$-phase to the $y$-phase and from the $y$-phase to the $xz$-phase at the vicinity of the transition point.
These calculations show the hysteresis, which is reasonable for the first order phase transition.
It is, however, noted that the hysteresis region of the field observed in our calculations is wider than usual, on which we will discuss later.

\subsection{Phenomenological field theory}
Using a Landau-Ginsburg type field theory, Zheludev~\textit{et al.}\cite{rf:K} have studied the transition between the Haldane phase and the ordering phase in NDMAP.
They succeeded in explaining some aspects of the transition.
We follow their method but extend their Lagrangian, so that it can describe the phase transition between the ordering phases:
we need more care about the anisotropy in higher order terms.
We start with the $S=1/2$ alternating chain Hamiltonian,\cite{rf:K}
\begin{equation}
  \label{eq:Ham-Hiwasa}
  \mathcal{H} = \sum_n \left[
    \sum_\alpha \left(
      J_\alpha S^\alpha_{2n-1} S^\alpha_{2n}
      + J'_\alpha S_{2n}^\alpha S_{2n+1}^\alpha
    \right)
    - \mathvec{h} \cdot \mathvec{S}_n \right],
\end{equation}
where $J_\alpha$, $J'_\alpha$ and \textvec{h} mean, respectively, two kinds of nearest-neighbor anisotropic exchange constants and the external magnetic field.
In comparison with the Hamiltonian they have used, this Hamiltonian is modified in the $J'$ term where the anisotropy is introduced.
The Lagrangian obtained by the similar procedures as described in ref.~\ref{rf:K}, is written as
\begin{equation}
  \label{eq:Leff}
  \mathcal{L} =
  \sum_{i=x,y,z} \left\{
    \frac{1}{\tilde{m}_i} \left[
      \left( \partial_t \phi_i \right)^2
      - v_i^2 \left( \partial_x \phi_i \right)^2
    \right]
    - 2 \frac{1}{\tilde{m}_i}
    \left( \mathvec{h} \times \mathvec{\phi} \right)_i \partial_t \phi_i
  \right\}
  - U_2 \left( \mathvec{\phi} \right)
  - U_4 \left( \mathvec{\phi}, \partial_t \mathvec{\phi} \right),
\end{equation}
where \textvec{\phi}, \textvec{v} and \textvec{h} denote, respectively, the staggered order parameter, which is related to the sublattice magnetization, the characteristic velocity and the external magnetic field.
The quadratic and quartic parts of the potential are given by
\begin{align}
  &U_2 \left( \mathvec{\phi} \right) =
  \sum_i \left[
    m_i \phi_i^2
    - \frac{1}{\tilde{m}_i} \left( \mathvec{h} \times \mathvec{\phi} \right)_i^2
  \right], \\
  \label{eq:U4-Hiwasa}
  &U_4\left( \mathvec{\phi}, \partial_t \mathvec{\phi} \right) =
  \sum_i \lambda_i \phi_i^2 \mathvec{\phi}^2
  + \sum_{ij} \lambda_{1,ij} \phi_i^2 \frac{1}{\tilde{m}_j^2} F_j^2
  + \sum_{ij} \lambda_{2,ij} \frac{\phi_i \phi_j}{\tilde{m}_i\tilde{m}_j} F_i F_j,\\
  &\hspace{1cm}\mathvec{F} =
  - \partial_t \mathvec{\phi}
  + \mathvec{h} \times \mathvec{\phi}.
\end{align}
We note that the modification appear only in $U_4$, which becomes more anisotropic.
If we set $\lambda_i=\lambda$, $\lambda_{1,ij}=\lambda_1$, $\lambda_{2,ij}=\lambda_2$, we obtain the quartic potential obtained by Zheludev \textit{et al.}\cite{rf:K}
Note that $\lambda_{2,ij}$ is a symmetric tensor, $\lambda_{2,ij} = \lambda_{2,ji}$, since $\partial_{\phi_i} (\partial_{\phi_j} U_4) = \partial_{\phi_j} (\partial_{\phi_i} U_4)$.

The transition from the Haldane phase to the ordering phase is suggested by a new static non-zero solution, $\mathvec{\phi}^{(0)}$, where the static part of the potential has a minimum.
This transition is characterized by the change of the sign of the second derivatives of $U_2$.
Two kinds of transitions, namely transitions from the Haldane phase to the $xz$-phase and to the $y$-phase, are possible depending on $\theta$.
The critical field $h_\mathrm{c}$ as a function of $\theta$ is obtained as
\begin{equation}
  h_\mathrm{c} = \min \left( h_\mathrm{c}^{(xz)}, h_\mathrm{c}^{(y)} \right),
\end{equation}
\begin{align}
  &h_\mathrm{c}^{(xz)}
  = \sqrt{ \frac{m_x \tilde{m}_y m_z }{ m_x \sin^2 \theta + m_z \cos^2 \theta} }, \\
  &h_\mathrm{c}^{(y)}
  = \sqrt { \frac{\tilde{m}_x m_y \tilde{m}_z }
    { \tilde{m}_x \sin^2 \theta + \tilde{m}_z \cos^2 \theta} },
\end{align}
where $h_\mathrm{c}^{(xz)}$ and $h_\mathrm{c}^{(y)}$ are the critical fields for two transitions.
Note that $h_\mathrm{c}^{(xz)}$ and $h_\mathrm{c}^{(y)}$ are independent from the modification of our Lagrangian since they depend only on $U_2$.

The static solution $\mathvec{\phi}^{(0)}$ can be obtained from the following simultaneous equation:
\begin{equation}
  \sum_j \Omega_{ij} \phi_j
  + \sum_{lmn} \Lambda_{il,mn} \phi_l \phi_m \phi_n = 0,
\end{equation}
where the matrices $\Omega$, $\Lambda$ are defined as
\begin{align}
  &\Omega_{ij} =
  m_i \delta_{ij}
  - \sum_{ll'n} \epsilon_{iln} \epsilon_{jl'n} \frac{h_l h_{l'}}{\tilde{m}_n}, \\
  &\Lambda_{ij,mn} =
  \Gamma_{ij,mn} + \Gamma_{mn,ij}, \\
  \Gamma_{ij,mn} =
  \lambda_i \delta_{ij} \delta_{mn}
  &+ \sum_{kk'll'} \left(
    \lambda_{1,ik} \delta_{ij} \delta_{kk'} + \lambda_{2,ij} \delta_{ik} \delta_{jk'}
  \right)
  \epsilon_{kln} \epsilon_{k'l'm} \frac{h_l h_{l'}}{\tilde{m}_k \tilde{m}_{k'}}.
\end{align}
Here $\epsilon_{ijk}$ means the Levi-Civita symbol.

The excitation energies are obtained by the equation of motion, where the small fluctuations around the static solution are treated up to quadratic order.
There are three excitation modes, of which energies are obtained as three real roots of the following secular equation:
\begin{equation}
  \det \left(
    \mathrm{M} - \omega^2 \mathrm{G} - i \omega \mathrm{C}
  \right)=0,
\end{equation}
where M, G, C are defined by
\begin{align}
  &\mathrm{M}_{ij} =
  \Omega_{ij} + \frac{\left( v_i q \right)^2}{\tilde{m}_i} \delta_{ij}
  + \sum_{mn} \left( \Lambda_{ij,mn} + \Lambda_{im,jn} + \Lambda_{in,mj} \right)
  \phi^{(0)}_m \phi^{(0)}_n, \\
  &\mathrm{G}_{ij} =
  \left[
    \frac{1}{\tilde{m_i}} - \sum_l \lambda_{1,li} \left( \frac{\phi^{(0)}_l}{\tilde{m}_i} \right)^2
  \right] \delta_{ij}
  - \lambda_{2,ij} \frac{\phi^{(0)}_i \phi^{(0)}_j}{\tilde{m}_i \tilde{m}_j}, \\
  &\mathrm{C}_{ij} = \mathrm{R}_{ij} - \mathrm{R}_{ji}, \\
  \begin{split}
    &\hspace{0.7cm}
    \mathrm{R}_{ij} =
    \frac{1}{\tilde{m}_i} \sum_l \epsilon_{ijl} h_l
    + \sum_{kk'l} \left(
      2 \lambda_{1,k'i} \epsilon_{ilk} \delta_{jk'}
      + \lambda_{1,k'j} \epsilon_{ilj} \delta_{kk'} \right)
    \frac{1}{\tilde{m}_i^2} h_l \phi^{(0)}_k \phi^{(0)}_{k'} \\
    &\hspace{1.7cm}
    + \frac{\phi^{(0)}_i}{\tilde{m}_i} \sum_{kl}
    \left( \lambda_{2,ij} \frac{1}{\tilde{m}_j} - \lambda_{2,ik} \frac{1}{\tilde{m}_k} \right)
    \epsilon_{jlk} \phi^{(0)}_k h_l .
  \end{split}
\end{align}

\begin{figure}[t]
  {\small (i) $\mathvec{h} \parallel c$ axis \hspace{5.75cm}
    (ii) $\mathvec{h} \parallel a$ axis}\\
  \includegraphics[width=0.495\textwidth]{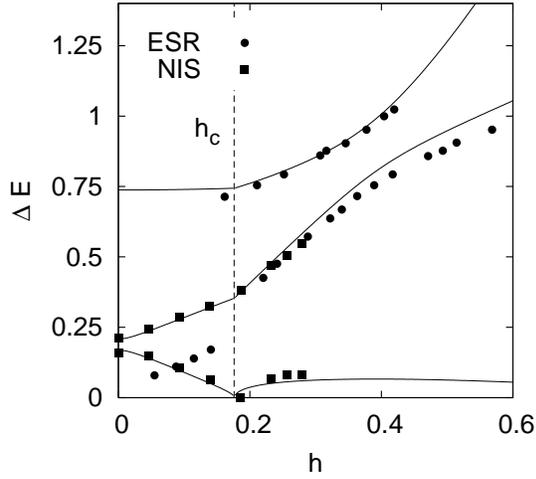}
  \includegraphics[width=0.495\textwidth]{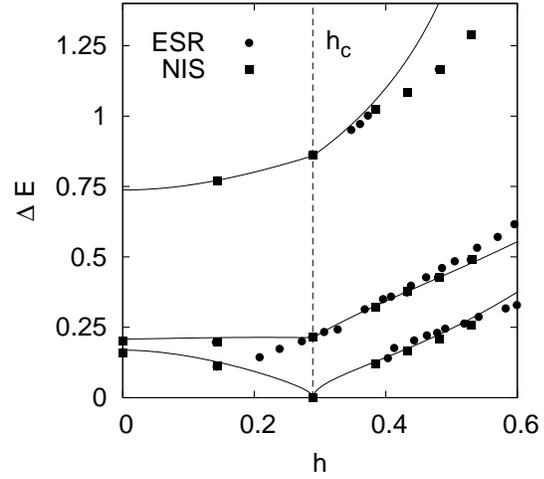}
  \caption{
    The lowest three excitation energies as a function of the magnetic field along (i) the $c$ axis and (ii) the $a$ axis.
    The circles and the squares denote the results observed by ESR\cite{rf:HHKKM} and by NIS\cite{rf:ZHBKSKPQ} with the wave vector $q=\pi$, respectively.
    The result calculated by PFT is shown by the solid lines.
    The vertical dashed lines denote the critical field $h_\mathrm{c}$, which is the boundary between the Haldane phase and the ordering phase.}
  \label{fig:expfit}
\end{figure}

The phenomenological parameters, $m, \tilde{m}, \lambda, \lambda_1$, and $\lambda_2$, are chosen so that PFT reproduces the magnetic field dependence of the excitation energies observed by the neutron inelastic scattering (NIS) and by the electron spin resonance (ESR) experiments under the magnetic fields along the $a$ and the $c$ axes.
Note that the case in which the magnetic field along the $a$ ($c$) axis corresponds to the case $\theta=0.6\pi$ ($\theta=0.1\pi$) for a spin chain and $\theta=0.4\pi$ ($\theta=-0.1\pi$) for another chain in NDMAP.
We determine six parameters, $m$ and $\tilde{m}$, from the data of the energy gaps and of the critical fields in the Haldane phase.
Remaining parameters, $\lambda$, $\lambda_1$ and $\lambda_2$, are chosen so that PFT reproduces the excitation energies in the ordering phase ($h>h_\mathrm{c}$).
Although these parameters can not be determined uniquely from NIS and ESR results, we chose a following set of the parameters with the aid of the result by DMRG, which will be shown soon later:
\begin{equation}
  \begin{array}{ccc}
    m_x=0.19,&m_y=0.29,&m_z=1.7\\
    \tilde{m}_x=0.15,&\tilde{m}_y=0.15,&\tilde{m}_z=0.32
  \end{array}
\end{equation}

\begin{equation*}
  \mathvec{\lambda}=
  \left( 1.4\ \ 1.0\ \ 1.0 \right),
\end{equation*}
\begin{equation}
  \mathvec{\lambda}_1=
  \left(
    \begin{array}{ccc}
      0.37&0.37&0.37\\
      0.37&0.37&0.25\\
      0.37&0.48&0.37
    \end{array}
  \right),\quad
  \mathvec{\lambda}_2=
  \left(
    \begin{array}{ccc}
      0.30&0.30&0.49\\
      0.30&0.30&0.44\\
      0.49&0.44&0.30
    \end{array}
  \right)
\end{equation}
In Fig.~\ref{fig:expfit}, we show the magnetic field dependence of the excitation energies calculated by PFT, which are compared with the experimental results obtained by ESR\cite{rf:HHKKM} and by NIS.\cite{rf:ZHBKSKPQ}
Our PFT can reproduce quite well the experimental results.

\section{Results}
\subsection{Energy Gap and Sublattice Magnetization}
\begin{figure}[t]
  {\small (i) $\theta=0$}\\
  \includegraphics[width=0.495\textwidth]{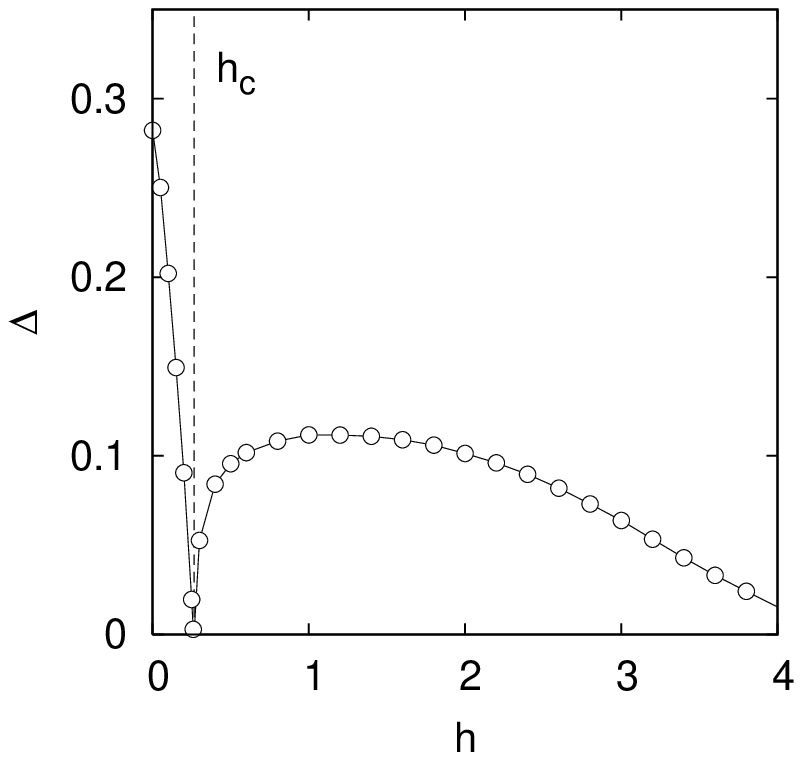}
  \includegraphics[width=0.495\textwidth]{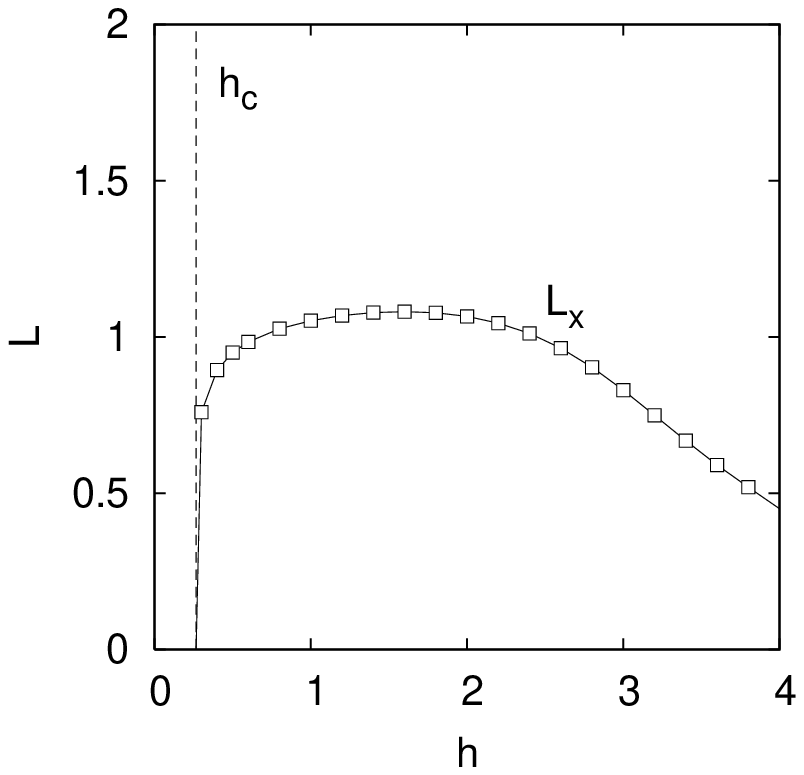}

  {\small (ii) $\theta=\pi/2$}\\
  \includegraphics[width=0.495\textwidth]{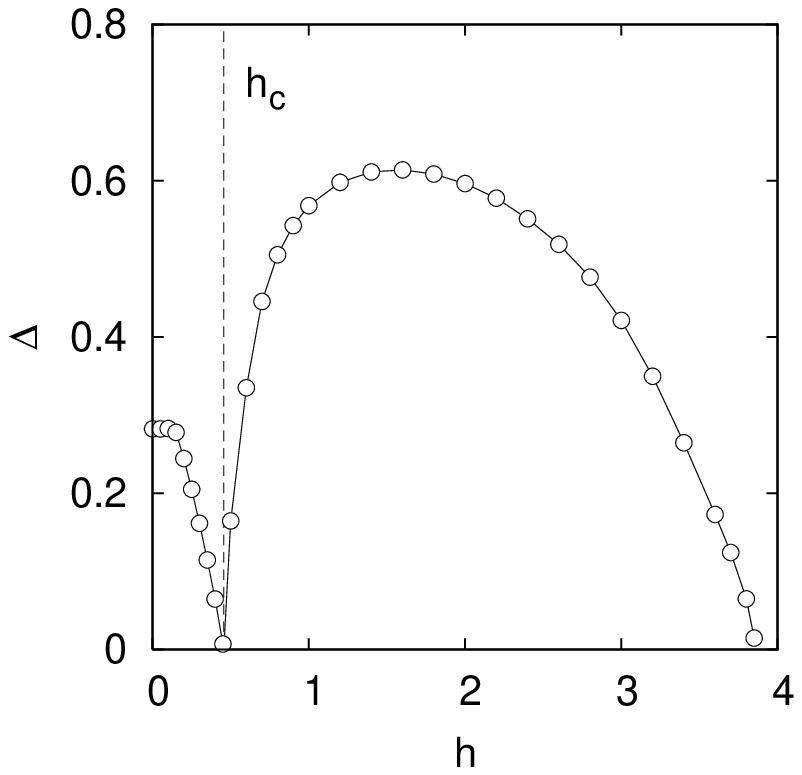}
  \includegraphics[width=0.495\textwidth]{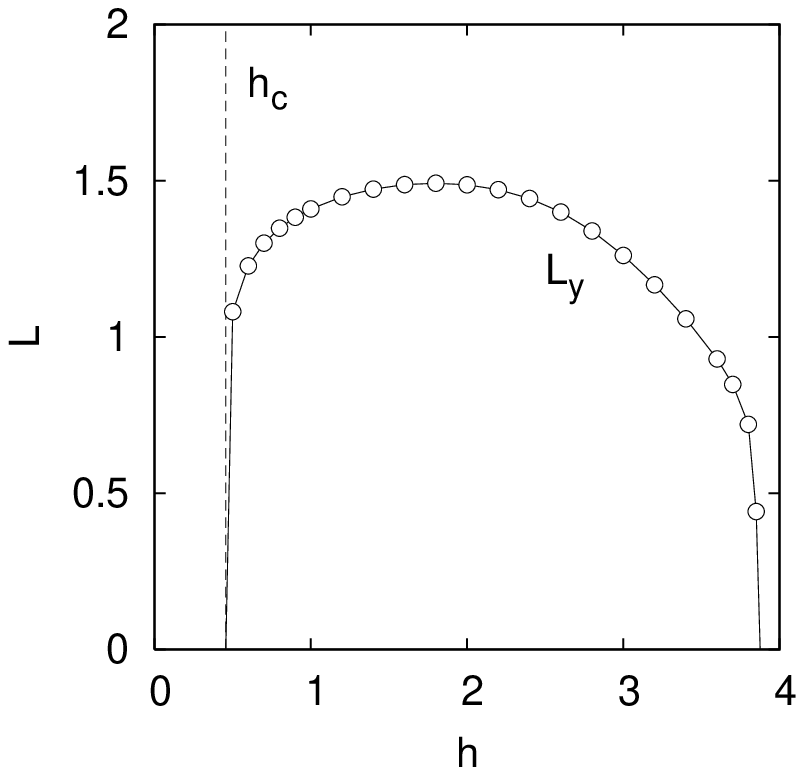}
  \caption{
    The energy gap $\Delta$ and the sublattice magnetization \textvec{L} as a function of $h$ at (i) $\theta=0$ and (ii) $\theta=\pi/2$ calculated by DMRG.
    The calculation has been done keeping up to 45 states ($m=45$).
    The Haldane gap closes in the case (i) at $h_\mathrm{c}=0.26$, and in the case (ii) at $h_\mathrm{c}=0.45$, and a gap open again in the ordering phase.
    The field-induced ordering phase is the $x$-phase for $\theta=0$, and the $y$-phase for $\theta=\pi/2$.}
  \label{fig:t0t5-dmrg}
\end{figure}

The energy gap $\Delta$ and the sublattice magnetization \textvec{L} in each phase are calculated as a function of the magnitude $h$ and the angle from the $z$ axis $\theta$ of the magnetic field, using DMRG as well as PFT.
The result calculated by DMRG as a function of $h$ at the fixed angles, $\theta=0$ and $\theta=\pi/2$, is shown in Fig.~\ref{fig:t0t5-dmrg}.
In both cases, the staggered ordering state appears above $h_\mathrm{c}$.
The direction of the sublattice magnetization, however, is different, depending on $\theta$:
\textvec{L} is parallel to the $x$ axis for $\theta=0$, while it is parallel to the $y$ axis for $\theta=\pi/2$.
The latter case corresponds to the situation of the spin-flop transition in the classical spin system but, in our quantum case, it is not difficult to realize that there is no quantum analog since the induced staggered magnetization is already perpendicular to the field.

\begin{figure}[t]
  {\small (i) $\theta=0$}\\
  \includegraphics[width=0.495\textwidth]{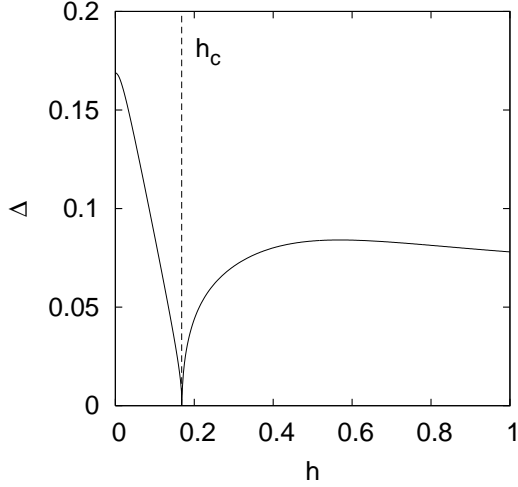}
  \includegraphics[width=0.495\textwidth]{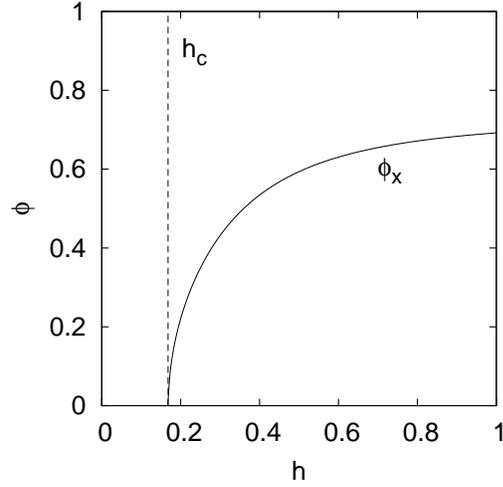}

  {\small (ii) $\theta=\pi/2$}\\
  \includegraphics[width=0.495\textwidth]{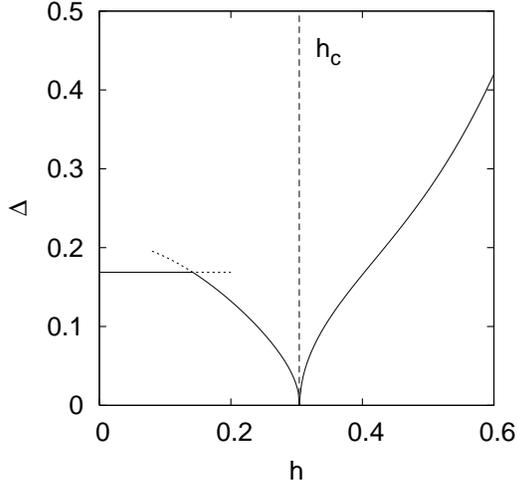}
  \includegraphics[width=0.495\textwidth]{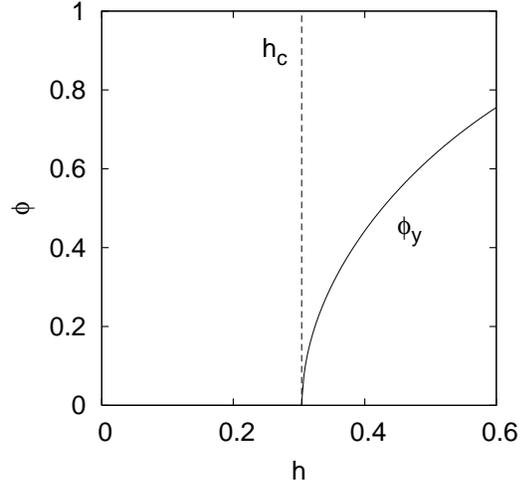}
  \caption{
    The energy gap $\Delta$ and the order parameter \textvec{\phi} as a function of $h$ at (i) $\theta=0$ and (ii) $\theta=\pi/2$ calculated by PFT.
    The Haldane gap closes in the case (i) at $h_\mathrm{c}=0.17$, and in the case (ii) at $h_\mathrm{c}=0.30$.
    Then, the field-induced ordering phase occurs.
    The orientation of the order parameter corresponds to the sublattice magnetization in DMRG;
    the $x$-phase for $\theta=0$ and the $y$-phase for $\theta=\pi/2$}
  \label{fig:t0t5-pft}
\end{figure}

In Fig.~\ref{fig:t0t5-pft}, we show the result calculated by PFT.
Because our Lagrangian describes behavior of the system only in the small values of \textvec{\phi}, we show the result up to $h=1$.
The order parameter is parallel to the $x$ axis for $\theta=0$ and to the $y$ axis for $\theta=\pi/2$, which is qualitatively the same as the sublattice magnetization obtained by DMRG in Fig.~\ref{fig:t0t5-dmrg}.

Next, let us consider the cases of intermediate angles.
We show $\Delta$ and \textvec{L} (\textvec{\phi}) calculated by DMRG and by PFT for $\theta=0.1\pi$, whose situation corresponds to the case where the magnetic field is applied along the $c$ axis in NDMAP.
The results obtained by DMRG and by PFT are qualitatively consistent:
with increasing the magnetic field, the Haldane gap becomes smaller and closes at $h_\mathrm{c}$, where the quantum phase transition occurs from the Haldane phase to the $xz$-phase similar to the case $\theta=0$.
It is surprising to find a successive phase transition:
the $xz$-phase further changes to the $y$-phase at the higher critical field, $h_\mathrm{sr}$.
Since this transition is first order, the $xz$-phase coexists with the $y$-phase in the field region $h_\mathrm{sr1}<h<h_\mathrm{sr2}$, where $h_\mathrm{sr1}$ ($h_\mathrm{sr2}$) is the field at which the $y$-phase (the $xz$-phase) becomes unstable.
The latter phase transition between two kinds of the ordering phases can be said to be ``spin-reorientation transition''.
\begin{figure}[t]
  {\small (a) DMRG}\\
  \includegraphics[width=0.495\textwidth]{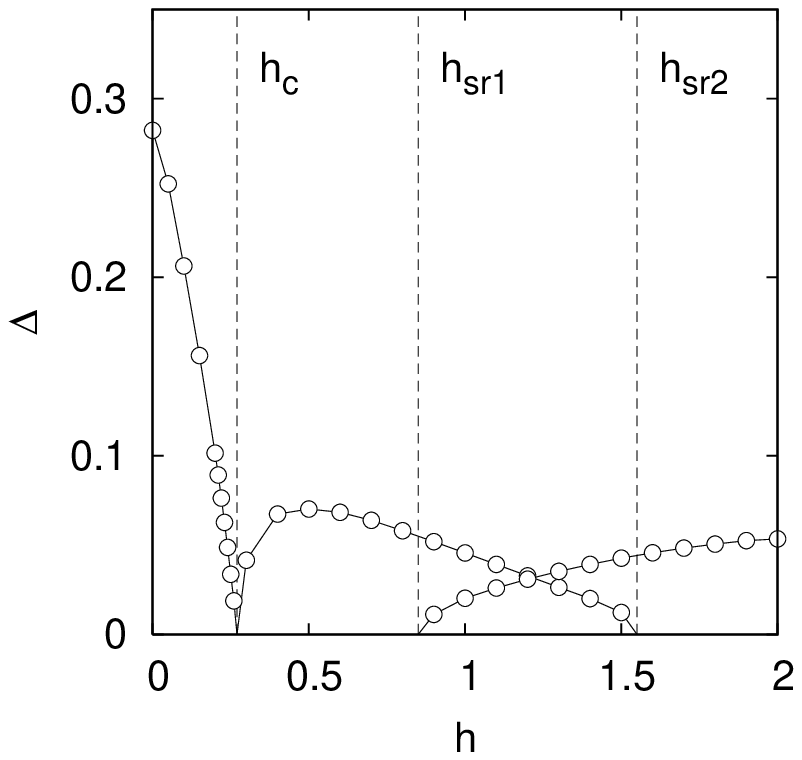}
  \includegraphics[width=0.495\textwidth]{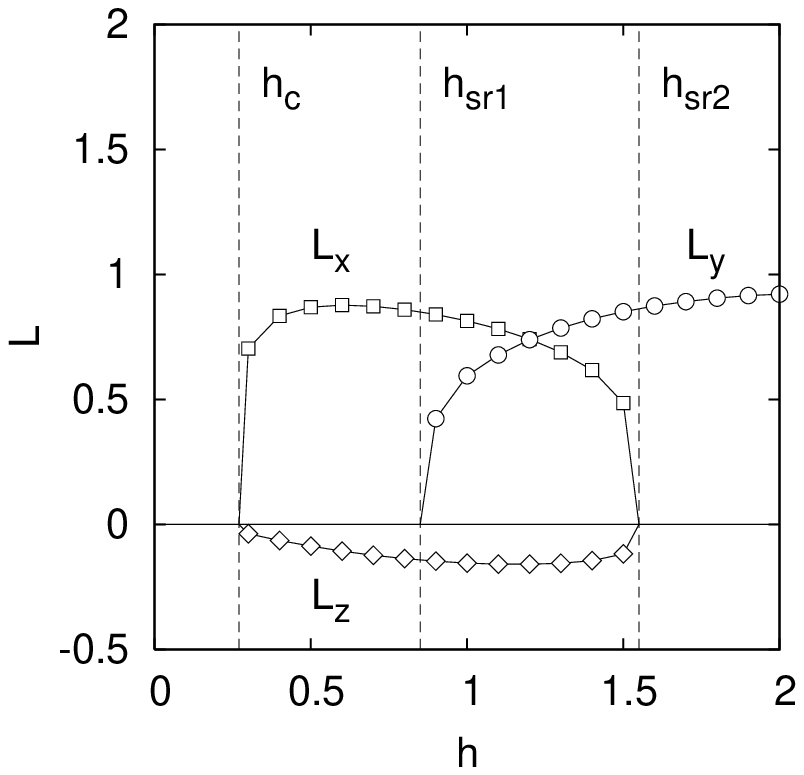}

  {\small (b) PFT}\\
  \includegraphics[width=0.495\textwidth]{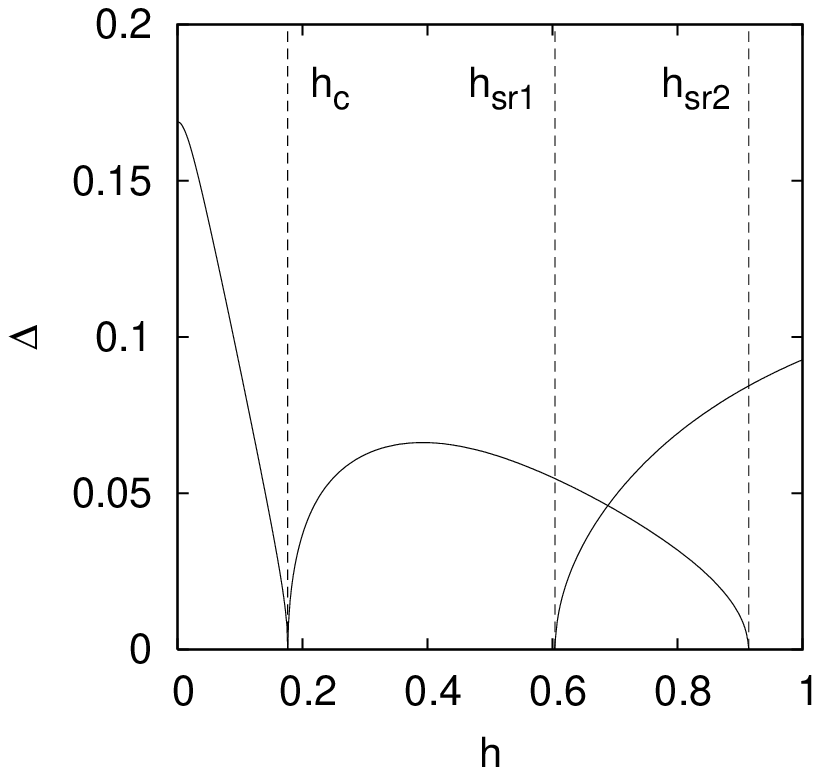}
  \includegraphics[width=0.495\textwidth]{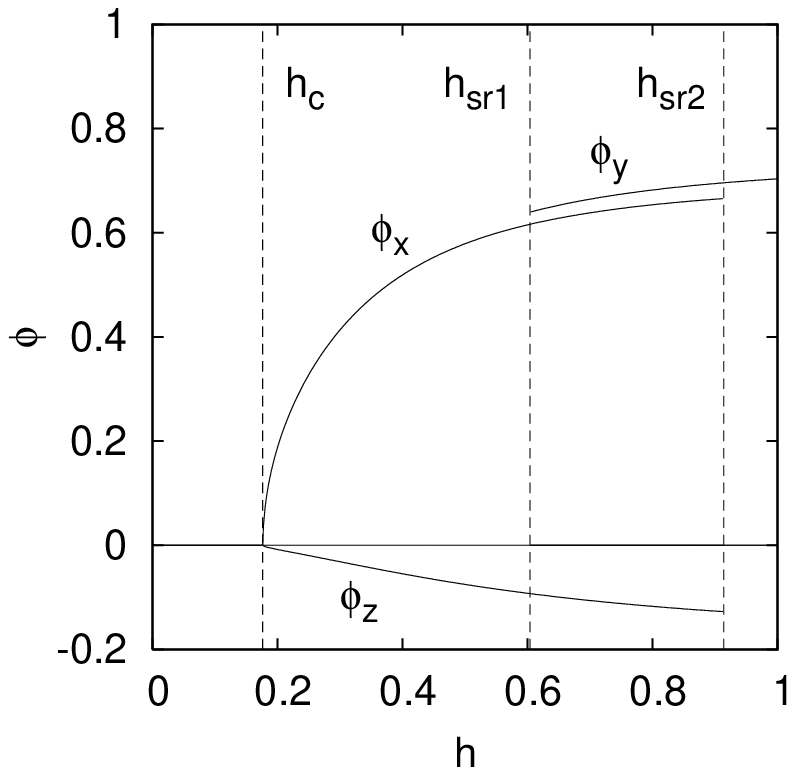}
  \caption{
    The energy gap $\Delta$ and the sublattice magnetization \textvec{L} (\textvec{\phi}) as a function of $h$ at $\theta=0.1\pi$ calculated by DMRG and by PFT.
    Two successive phase transitions can be seen in these figures:
    the one is the order-disorder phase transition at $h_\mathrm{c}$, and the other is the first order phase transition from the $xz$-phase to the $y$-phase: the spin-reorientation transition.
    The $xz$-phase and the $y$-phase become unstable, respectively, at $h_\mathrm{sr2}$ and at $h_\mathrm{sr1}$.}
  \label{fig:t1}
\end{figure}

\begin{figure}[t]
  {\small (a) DMRG}\\
  \includegraphics[width=0.495\textwidth]{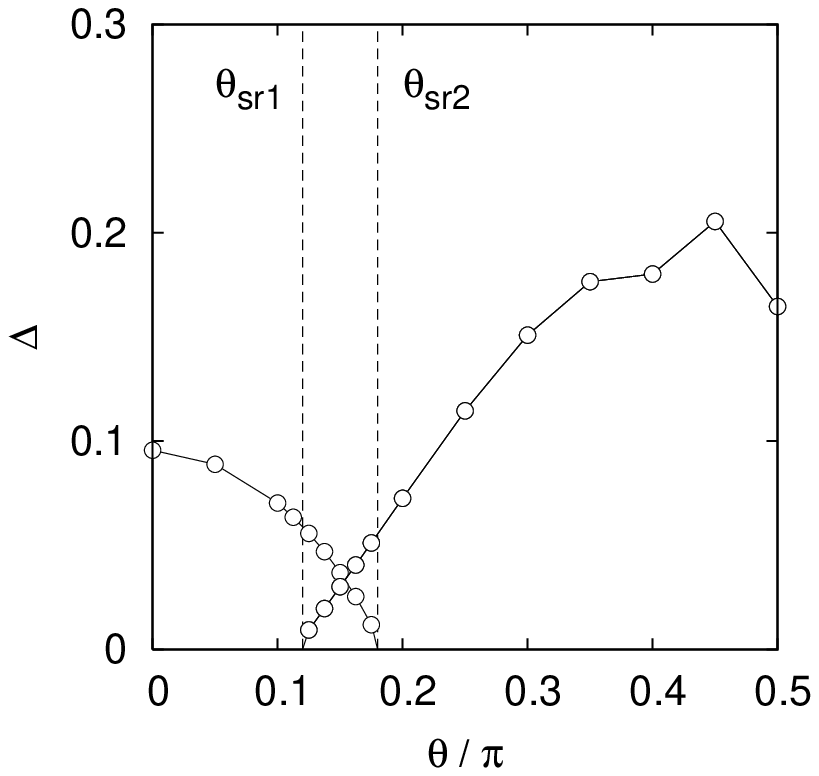}
  \includegraphics[width=0.495\textwidth]{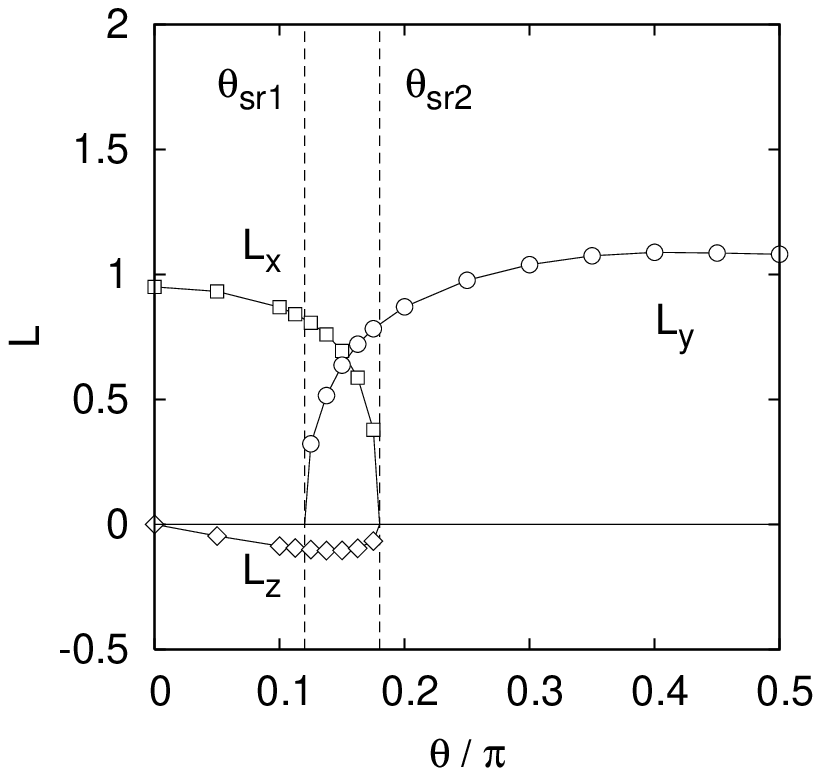}

  {\small (b) PFT}\\
  \includegraphics[width=0.495\textwidth]{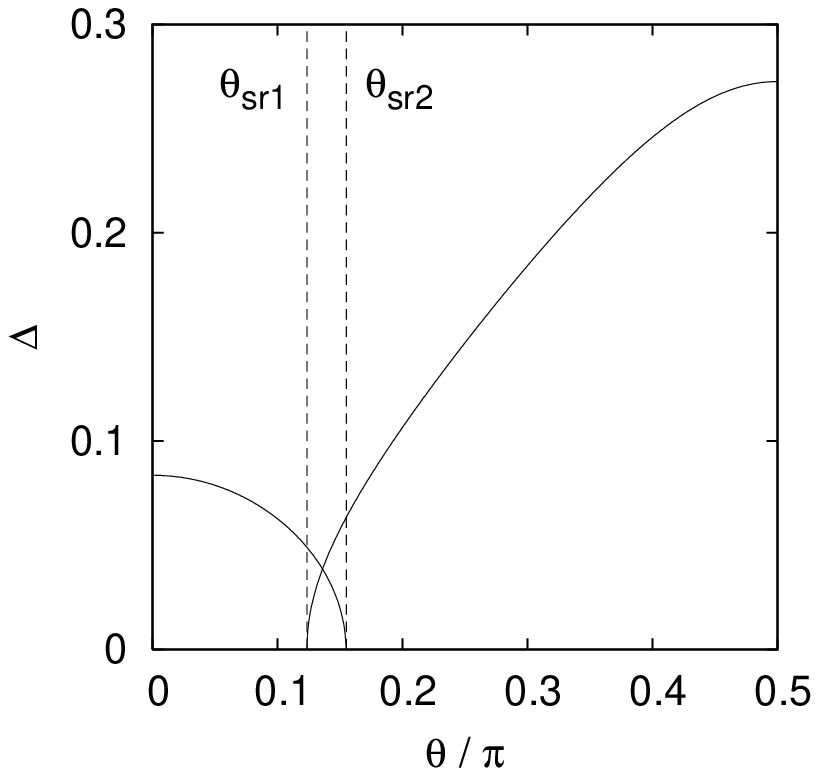}
  \includegraphics[width=0.495\textwidth]{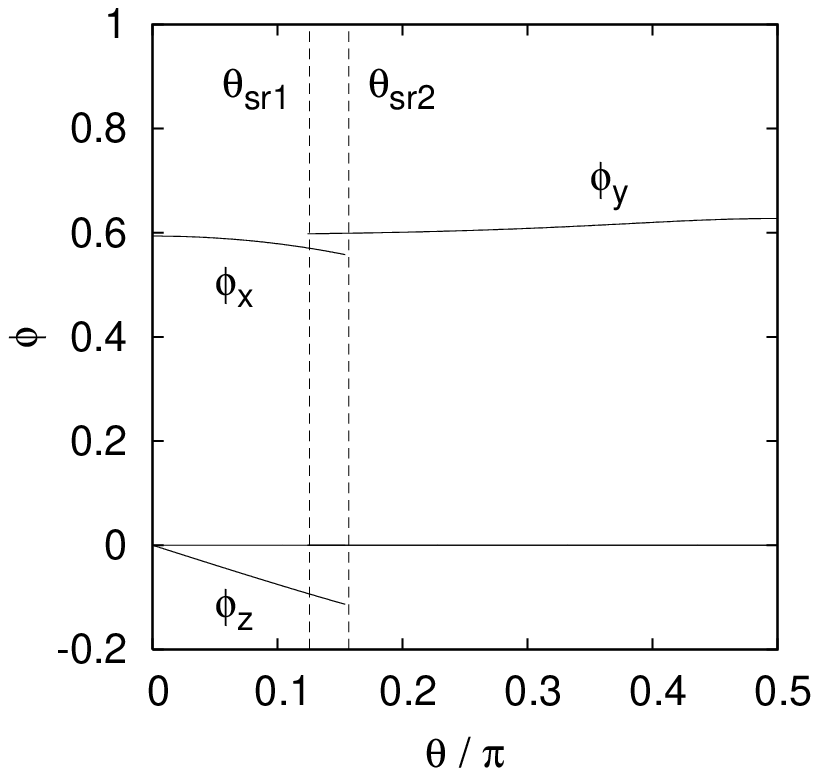}
  \caption{
    The energy gap $\Delta$ and the sublattice magnetization \textvec{L} (\textvec{\phi}) as a function of the field angle $\theta$ at $h=0.5$ by DMRG and by PFT.
    The $xz$-phase and the $y$-phase become unstable, respectively, at $\theta_\mathrm{sr2}$ and at $\theta_\mathrm{sr1}$.}
  \label{fig:h5}
\end{figure}

Next, we show $\Delta$ and \textvec{L} (\textvec{\phi}) as a function of $\theta$, in order to investigate the spin-reorientation transition.
In Fig.~\ref{fig:h5}, we see that, with increasing the field angle $\theta$, the $xz$-phase becomes unstable against the $y$-phase.
At the intermediate angle, $\theta_\mathrm{sr1}<\theta<\theta_\mathrm{sr2}$, both phases coexist, which is a typical signature of the first order phase transitions.

\subsection{Phase Diagram in the $h$-$\theta$ plane}
\begin{figure}[t]
  {\small (a) DMRG \hspace{0.35\textwidth} (b) PFT}\\
  \includegraphics[width=0.495\textwidth]{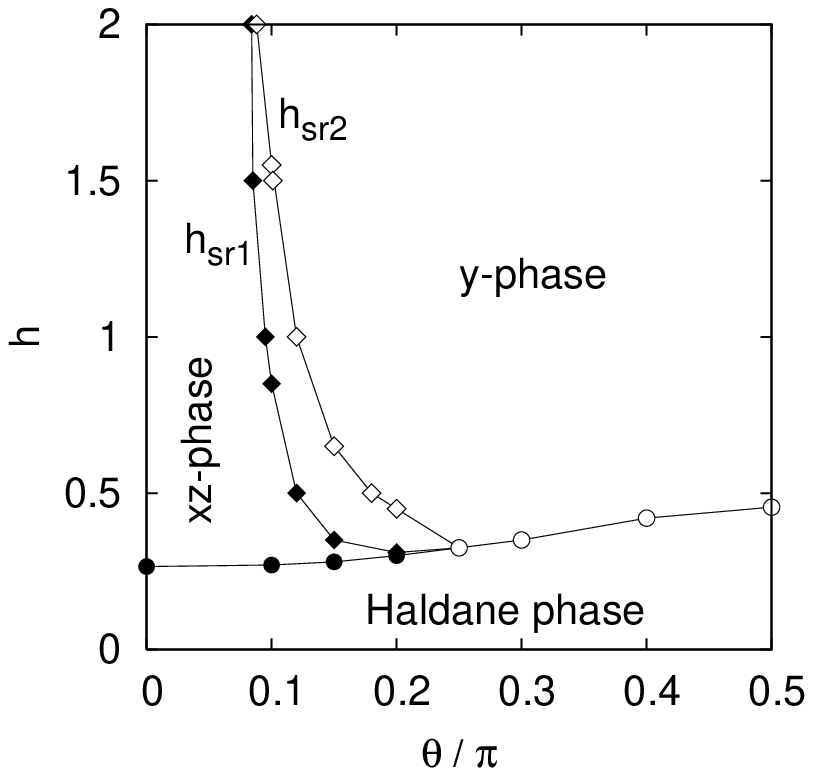}
  \includegraphics[width=0.495\textwidth]{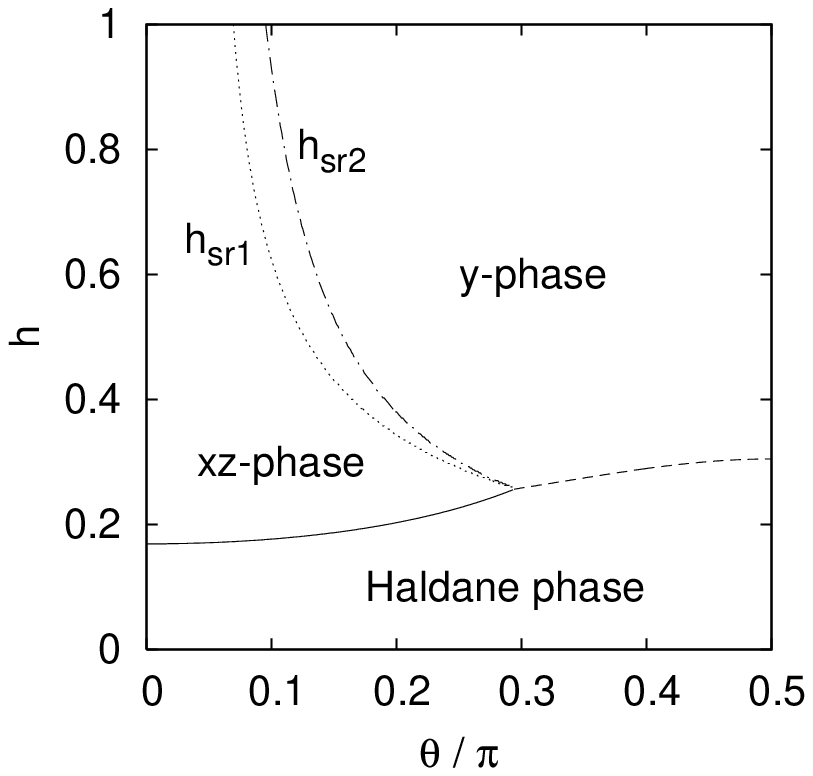}
  \caption{
    The phase diagram in the $h$-$\theta$ plane obtained (a) by DMRG and (b) by PFT.
    In (a), the closed and the open circles represent the critical fields, respectively, from the Haldane phase to the $xz$-phase and from the Haldane phase to the $y$-phase, while the closed and the open diamonds denote, respectively, the points, where the $xz$-phase and the $y$-phase become unstable.
    The lines are guides for the eye. In (b), the solid and the dashed curves correspond, respectively, to the closed and the open circles in (a), while the dotted and the chain curves correspond, respectively, to the solid and the open diamonds.}
  \label{fig:pd}
\end{figure}

We calculate the energy gap and the sublattice magnetization with respect to $h$ and $\theta$ by DMRG as well as by PFT.
From these results, we extract the phase diagram in the $h$-$\theta$ plane shown in Fig.~\ref{fig:pd}.
The phase diagrams obtained by DMRG and by PFT do not coincide with each other but are qualitatively similar:
These phase diagrams consist of the Haldane phase, the $xz$-phase and the $y$-phase.
It is interesting to note that when the applied magnetic field tilts from the $z$ axis, the spin-reorientation transition can occur at the higher critical field $h_\mathrm{sr}$ following the order-disorder phase transition at the lower critical field $h_\mathrm{c}$.

\section{Discussions}
We have studied the field-induced phase transition in the anisotropic $S=1$ Haldane chain and found the new phase transition at the higher critical field than that of the order-disorder phase transition by means of DMRG and of PFT.
Under the external magnetic field in the $x$-$z$ plane, depending on the field angle $\theta$, two kinds of the ordering phases are induced at $h_\mathrm{c}$, at which the Haldane gap closes:
when the magnetic field is applied along near the $z$ axis the $xz$-phase is induced, while it is along near the $x$ axis the $y$-phase is induced.
The spin-reorientation transition from the former to the latter further occur in a certain range of the angle $\theta$.
In the vicinity of the spin-reorientation transition field, the $xz$-phase and the $y$-phase coexist together and the magnetization process shows hysteresis, which is typical for the first order phase transition.

Although our result of DMRG keeping the states $m=45$ in the cycle at fields near the higher critical field does not converge so well with respect to $m$, it does in fields near the lower critical field.
Then, for a quantitative comparison of the high field transition with the experimental result, we need more sophisticated calculations, although at present we can not extend $m$ in our calculation because of the computational limitation.
On the other hand, PFT is flexible involving many parameters but it is phenomenological.
Nevertheless, we may expect that these methods provide a qualitatively correct description, considering the reasonable results obtained by both methods.

Considering these limitations, let us remind the experimental finding in NDMAP, a new phase transition at the high field.
The novel phase transition of the first order at high fields predicted by our theory is favorably compared with the phase transition observed, since in the experiment the magnetic field is applied along the $c$ axis ($\lbar \theta / \pi \rbar = 0.1$ for both sublattices) in NDMAP is located in the field region where the successive phase transition occurs.
We mention that this is a consequence of the tilting of the anisotropy axes from those of the crystal.
It is interesting to confirm experimentally the fact predicted: the critical field $h_\mathrm{sr}$ is very sensitive to the angle $\theta$ near $\theta / \pi \simeq 0.1$ as is seen in Fig.~\ref{fig:pd}.
Also NMR experiments to study the direction of the order parameter in each phase are desirable.
Another point we should note for the application of our theory to NDMAP is a choice of the anisotropy parameters $d=0.25$ and $e=0.01$.
These values, especially the latter, have not been determined so accurately yet, but it dominates the critical field of the spin-reorientation transition.
Then, in order to study this transition more quantitatively it needs to adjust these parameter values more carefully considering other experimental results.

In conclusion, with the aid of the results of DMRG as well as of PFT, we can draw the physics in the field-induced quantum ordering phases and that in the phase transition between them, which gives rise to the scenario to explain the experimental finding in NDMAP, the new phase transition at the high field.

\section*{Acknowledgements}
The authors would like to thank Dr. T. Sakai for his valuable discussions.
This work has been partially supported by Grant-in-Aid for Scientific Research on priority Areas ``High Field Spin Science in 100T'' (No.451) from the Ministry of Education, Culture, Sports, Science and Technology(MEXT)бе

\end{document}